\documentclass[11pt,twoside]{article}

\usepackage{asp2006}
\usepackage{epsf}
\usepackage{psfig}
\usepackage{lscape}
\usepackage{natbib}
\usepackage{graphicx}

\begin{document}
\title{Supersonically Turbulent, Shock Bound Interaction Zones}   

\author{Doris Folini and Rolf Walder} 
\affil{\'{E}cole Normale Sup\'{e}rieure, Lyon, CRAL, UMR CNRS 5574, 
       Universit\'{e} de Lyon, France} 

\author{Jean M. Favre} 
\affil{Swiss National Supercomputing Centre (CSCS), 
       CH-6928 Manno, Switzerland}    

\begin{abstract} 
Shock bound interaction zones (SBIZs) are ubiquitous in astrophysics.
We present numerical results for 2D and 3D, plane-parallel, infinitely
extended SBIZs. Isothermal settings and parameterized 
cooling are considered. We highlight and compare characteristic
of such zones. We emphasize the mutual coupling between the turbulence
within the SBIZ and the confining shocks, point out potential
differences to 3D periodic box studies of supersonic turbulence, and
contemplate on possible effects on the X-ray emission of such zones.
\end{abstract}


%
%
\section{Introduction}
\label{sec:intro}
Supersonically turbulent, shock-bound interaction zones (SBIZs) are
important for a variety of astrophysical objects. They contribute to structure formation in molecular
clouds~\citep{hunter-et-al:86, audit-hennebelle:05,
vazquez-semadeni-et-al:06, hennebelle-et-al:08} and to galaxy
formation~\citep{anninos-norman:96, kang-et-al:05}.  They affect the
X-ray emission of hot-star winds~\citep{owocki-et-al:88,
feldmeier-et-al:97} and the physics and
emitted spectrum of colliding wind binaries~\citep{stevens-et-al:92,
nussbaumer-walder:93, myasnikov-zhekov:98, folini-walder:00}. A large
number of numerical and theoretical studies have greatly improved
our understanding of SBIZs under various
conditions~\citep{vishniac:94, blondin-marks:96, walder-folini:98,
heitsch-et-al:05, folini-walder:06}

In this paper, we want to highlight and compare some of the
characteristics of such interaction zones under different conditions:
isothermal - radiatively cooling, 2D - 3D, symmetric - asymmetric
settings, early times - late times.

\section{Model Problem and Numerical Method}
\label{sec:model}
\begin{figure}[ht]
\begin{center}
\includegraphics[width=13cm,height=3.5cm]{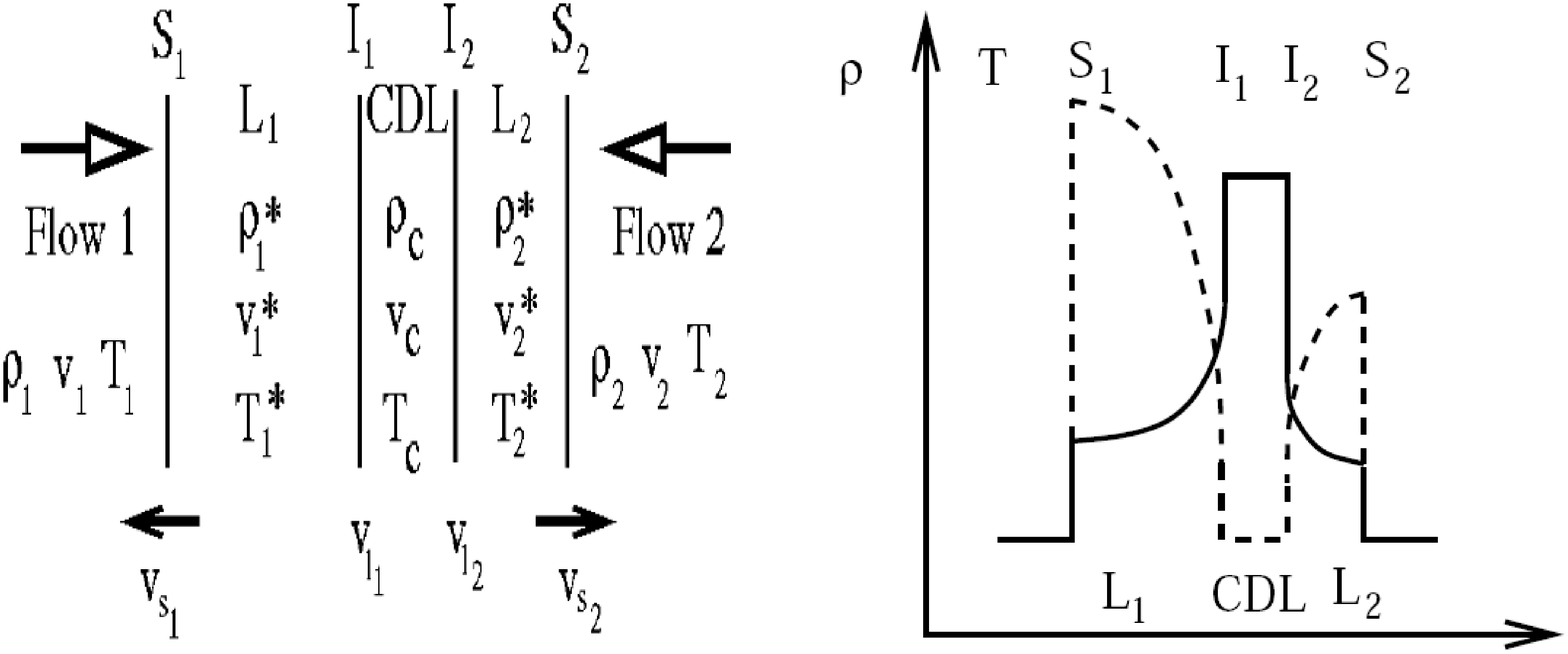}
\end{center}
\vspace{-0.5cm}
\caption{Sketch of the SBIZ. Radiatively cooled
         material piles up in the CDL. Two hot post
         shock layers (L1 and L2) exist in the case of parameterized
         cooling. Under isothermal conditions the SBIZ
         consists only of the CDL.}
\label{fig:sketch2d}
\end{figure}
We consider 2D and 3D, plane-parallel, infinitely extended
SBIZs. Both, isothermal and radiatively cooling settings are
investigated.  Two high Mach-number flows, oriented parallel (left
flow, subscript $l$) and anti-parallel (right flow, subscript $r$) to
the x-direction, collide head on. The resulting SBIZ is oriented in
the y-z-plane. It consists of a cold, dense layer (CDL) and, in the
case of radiative cooling, two layers of hot shocked gas
(Fig.~\ref{fig:sketch2d}). We investigated this system within the
frame of Euler equations. In the polytropic equation of state we take
$\gamma = 1.000001$ (isothermal) or $\gamma=5/3$ (radiatively
cooling). The right hand side of the energy equation we set to zero
(isothermal) or to a parameterized radiative loss
function~\citep{walder-folini:96}. Codes from the A-MAZE code
package~\citep{amaze:00} are used. A coarse mesh is used for the
upwind flows, a finer mesh for the CDL.  Typical refinement factors
are $2^{8}$ in 2D and $2^{4}$ in 3D, resulting in 1280 (2D) and 256
(3D) cells in the y-direction. The meshes adapt automatically to the
spatial extension of the CDL. The solution (density) on two adjacent
refinement levels is compared, cells where the difference exceeds a
prescribed tolerance are flagged. Refinement is applied to rectangular
blocks that contain all flagged cells and typically also some
unflagged cells.
\section{Characteristics of Interaction Zones}
\label{sec:characteristics}
\paragraph{Isothermal SBIZ:} Average
quantities, like slab thickness, auto-correlation length of
the confining shocks, or energy loss per unit volume in the
CDL, evolve approximately self-similarly for symmetric flow
collisions, where $\rho_{\mathrm{l}} = \rho_{\mathrm{r}}
\equiv \rho_{\mathrm{u}}$ and $M_{\mathrm{l}} = M_{\mathrm{r}}
\equiv M_{\mathrm{u}}$~\citep{folini-walder:06}. In 2D and 3D the root mean
square Mach number $M_{\mathrm{rms}}$ of the CDL scales linearly with
$M_{\mathrm{u}}$, the mean density $\rho_{\mathrm{m}}$ is independent
of $M_{\mathrm{u}}$. This contrasts with the 1D case, where 
$\rho_{\mathrm{m}} \propto M_{\mathrm{u}}^{2}$.
\begin{figure}[ht]
\begin{center}
\centerline{
\includegraphics[width=6cm]{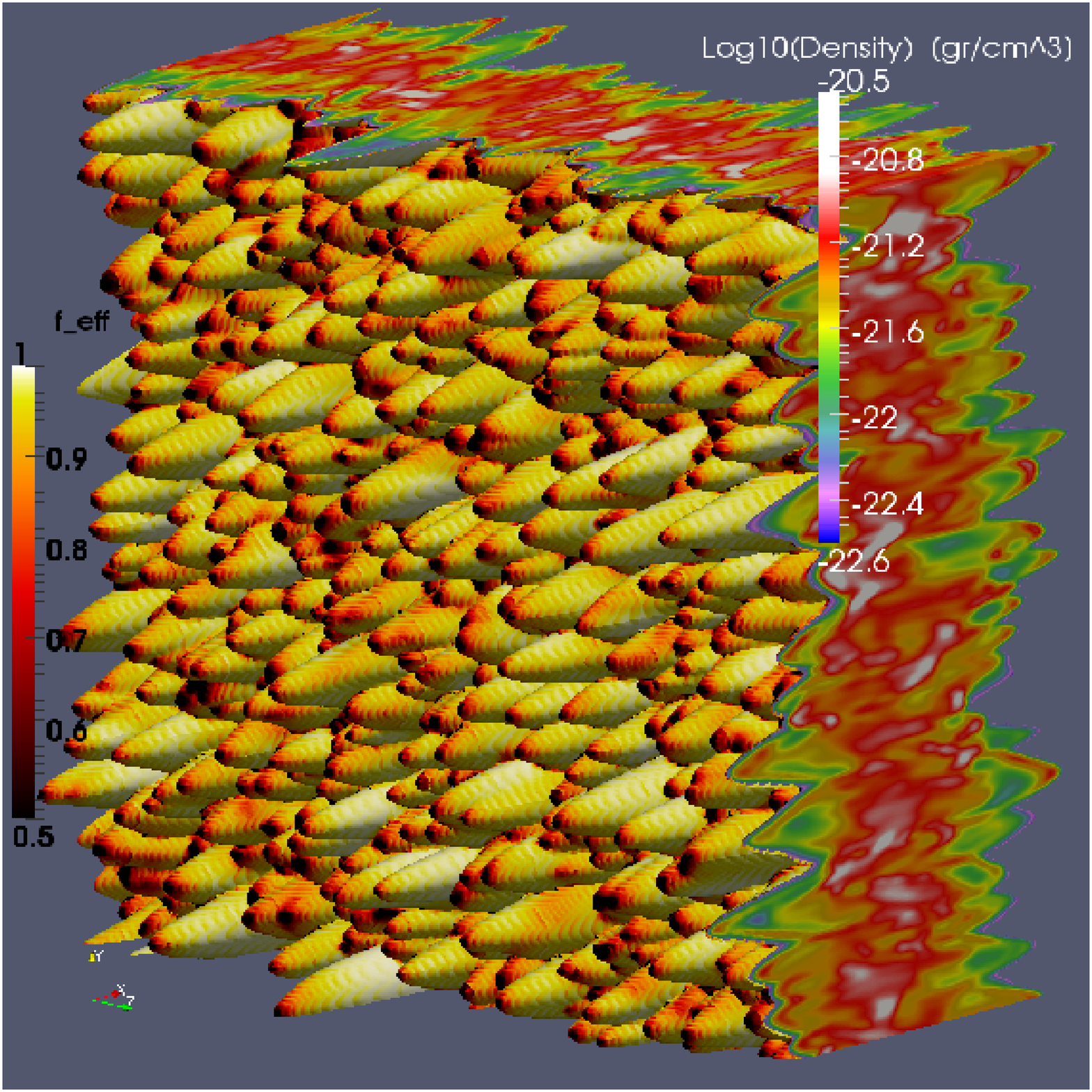}
\includegraphics[width=7cm,height=6cm]{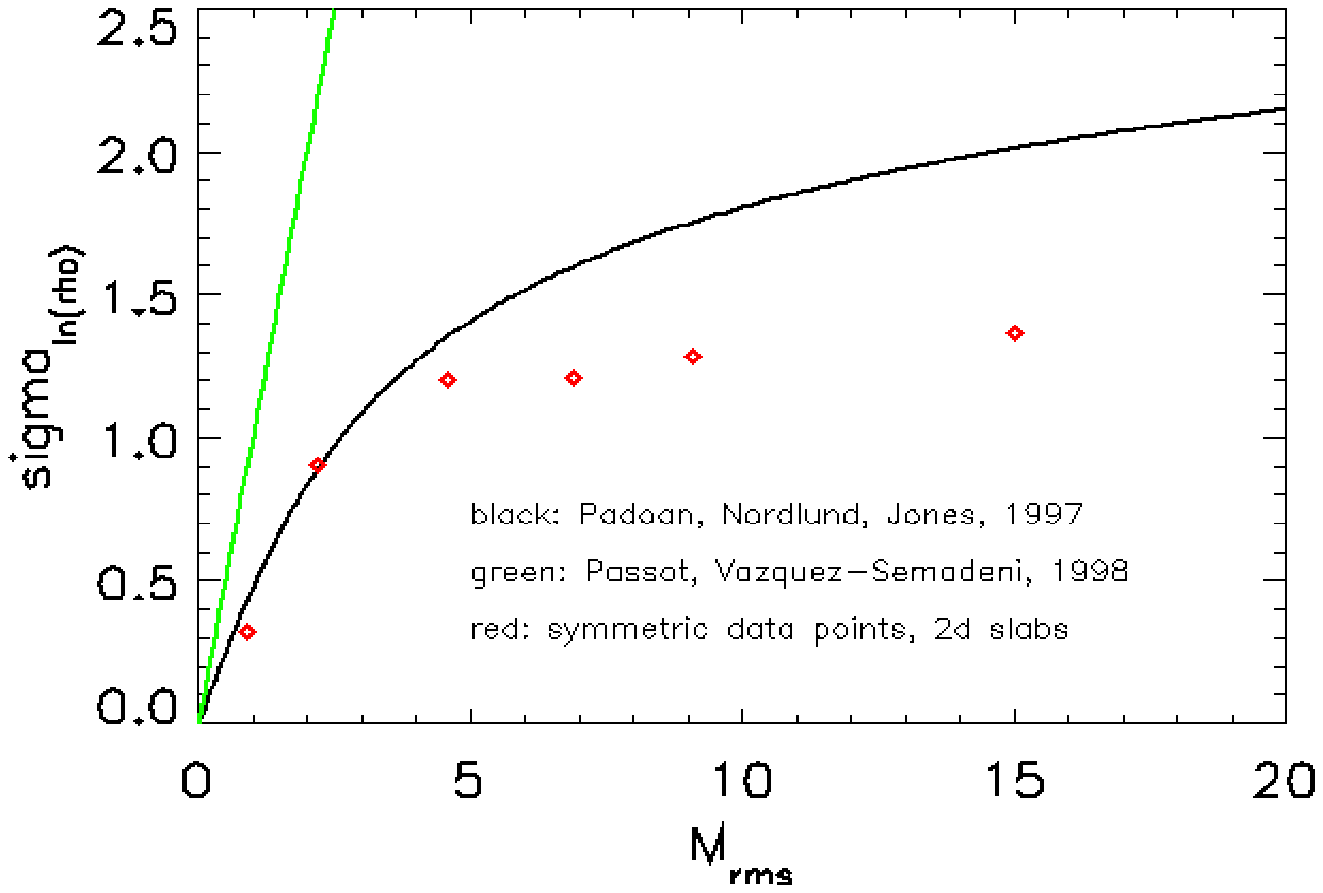}
}
\end{center}
\vspace{-0.8cm}
\caption{{\bf Left:} Isothermal SBIZ in 3D for $M_{\mathrm{u}} = 22$. 
         Shown is density, the confining shock is colored according to
         the local value of $f_{\mathrm{eff}}$. On average, the
         confining shocks are steeper in 3D ($f_{\mathrm{eff}} \approx
         0.83$ for $M_{\mathrm{u}} = 22$) than in 2D
         ($f_{\mathrm{eff}} \approx 0.59$ for $M_{\mathrm{u}} =
         22$). {\bf Right:} $\sigma(ln(\rho))$ versus
         $M_{\mathrm{rms}}$ from 2D symmetric slab simulations (red
         diamonds). The 2D slab results deviate from published fitting
         functions by~\citet{padoan-nordlund-jones:97} (black)
         and~\citet{passot-vazquez:98} (green) for large
         $M_{\mathrm{rms}}$.}
\label{fig:feff}
\end{figure}
\begin{figure}[ht]
\begin{center}
\centerline{
\includegraphics[width=7cm,height=4.5cm]{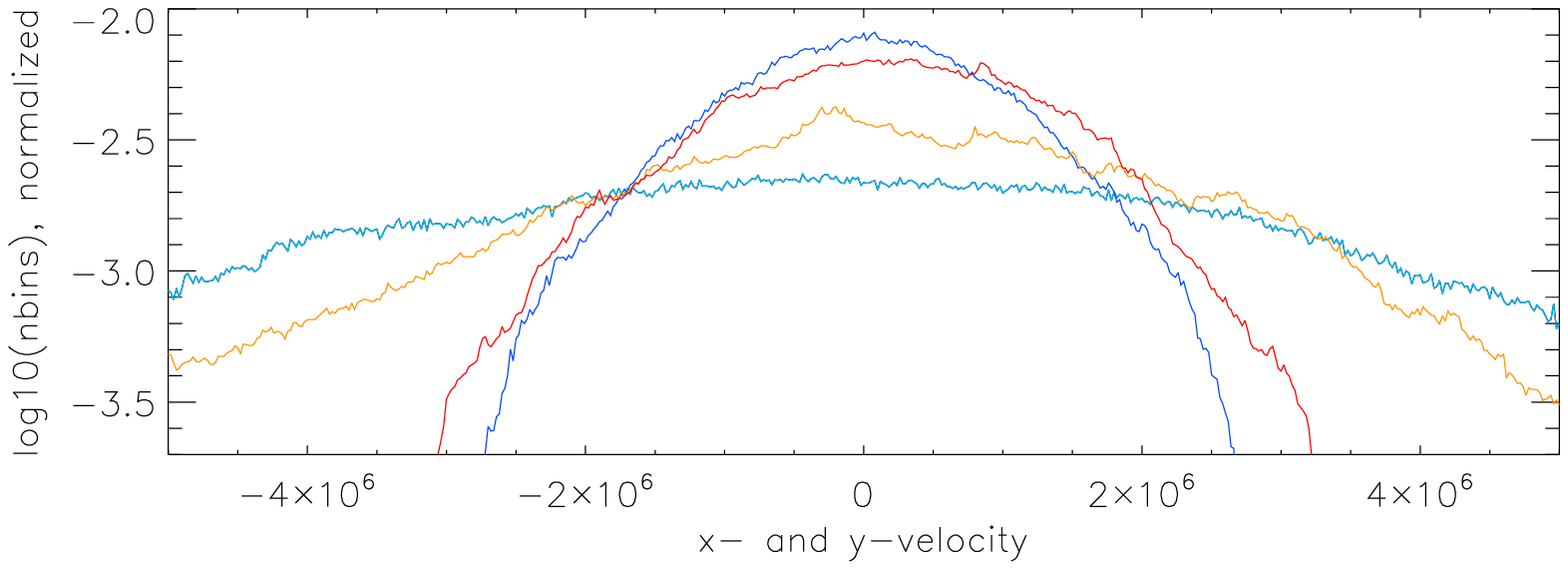}
\includegraphics[width=7cm,height=4.5cm]{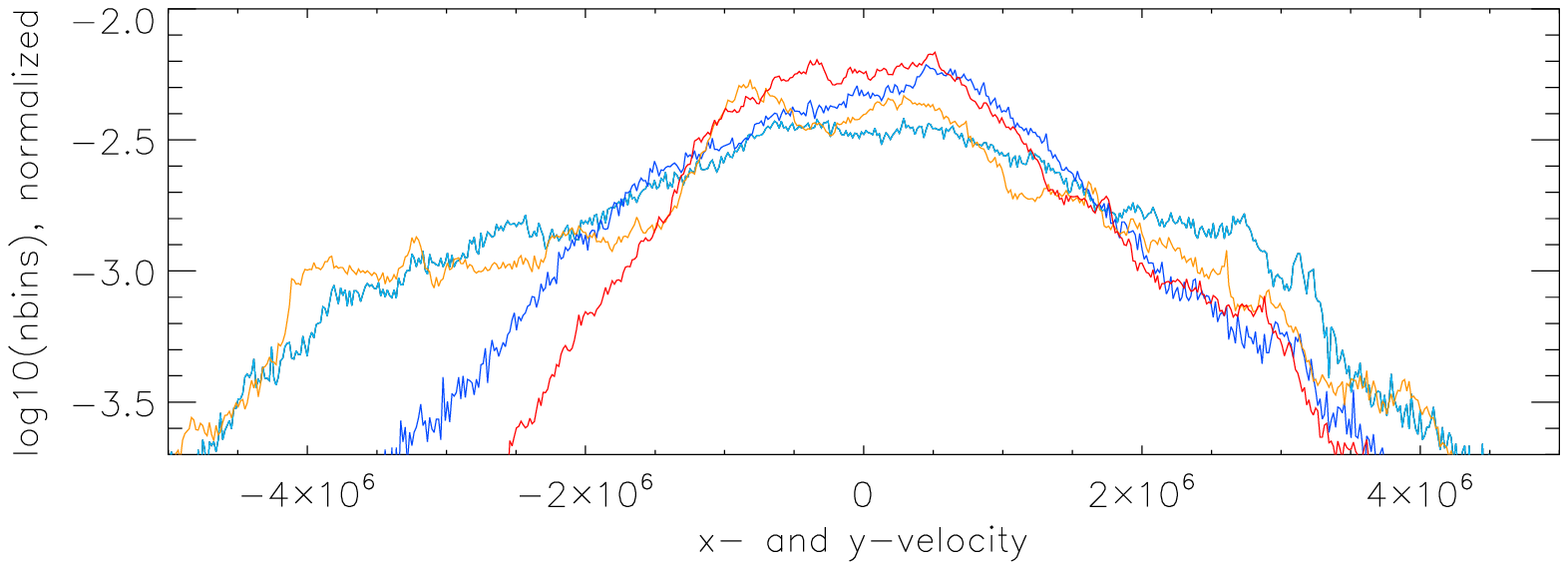}
}
\end{center}
\vspace{-1.4cm}
\caption{Velocity pdfs (x- and y-direction, two times) for an isothermal 
         (left) and a radiatively cooling (right) 2D simulation with
         $M_{\mathrm{u}}=22$ or $v_{\mathrm{u}} = 1.8 \cdot 10^{7}$
         cm/s. An early (bluish colors) and a late time (redish
         colors) are shown. Wider (narrower) pdfs represent
         x-velocities (y-velocities). Density plots of the cooling
         case are shown in Fig.~\ref{fig:visht} (left and right
         frame).}
\label{fig:velopdf}
\end{figure}

An intricate, 'self-regulating' interplay exists between the
turbulence within the CDL and its driving: $f_{\mathrm{eff}} = 1 -
M^{\beta}_{\mathrm{rms}}$ with $\beta
\approx -0.6$ in 2D. The fraction $f_{\mathrm{eff}}$ of the
upstream kinetic energy that passes the confining shocks unthermalized
and $M_{\mathrm{rms}}$ affect each other.  The larger
$M_{\mathrm{u}}$, the larger $M_{\mathrm{rms}}$, and the larger the
fraction of the upstream kinetic energy that is thermalized only {\it
within} the CDL. The dependence persists for asymmetric settings,
where $\rho_{\mathrm{l}} \ne \rho_{\mathrm{r}}$ and $M_{\mathrm{l}}
\ne M_{\mathrm{r}}$~\citep{folini-walder:06}. In 3D we find the confining shocks to be 
steeper than in 2D (Fig.~\ref{fig:feff}, left), thus
$f_{\mathrm{eff}}$ and $M_{\mathrm{rms}}$ are larger while
$\rho_{\mathrm{m}}$ is smaller.

The CDL is patchy. Patch sizes increase with the x-extension of the
CDL and with decreasing $M_{\mathrm{u}}$, as does the auto-correlation
length of the confining shocks. The density variance follows
$\sigma^{2}(ln(\rho)) = ln (1 +
M^{2}_{\mathrm{rms}}/4)$~\citep{padoan-nordlund-jones:97} up to
$M_{\mathrm{rms}} \approx 5$ but then levels off (Fig.~\ref{fig:feff},
right). This may be related to the strongly anisotropic velocity field
within the CDL, with larger (smaller) velocities along (perpendicular
to) the upstream flow direction (Fig.~\ref{fig:velopdf}, left).

\paragraph{One sided cooling SBIZ:}  Astrophysical examples include
wind blown bubbles or planetary nebula, where a fast wind collides with
a slow precursor, forming a hot ($>10^{7}$ K) reverse and a cool
($<10^{6}$ K) forward shock. A characteristic feature of SBIZs where
one of the confining shocks cools efficiently but not the other are
clumps of cold material that break off from the CDL and drift into the
non-cooling, hot post-shock material~\citep{walder-folini:98}. Density
images show a striking similarity with HST pictures of the
Helix nebula. A prerequisite is that the corresponding
boundary of the CDL is occasionally hit by a shock wave, which
triggers a Richtmeyer-Meshkov instability. The shock wave can be
generated by a thermal instability of the other hot post shock
layer~\citep{strickland-blondin:95, walder-folini:96} or by
inhomogeneities in the upstream flow. The CDL is weakly turbulent, the
velocity field is isotropic.

\begin{figure}[ht]
\begin{center}
\includegraphics[width=13cm,height=5.5cm]{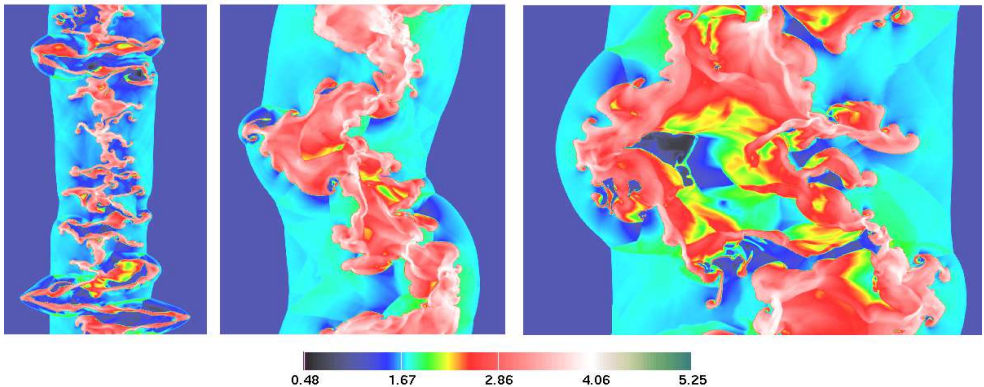}
\end{center}
\vspace{-0.8cm}
\caption{Two sided cooling SBIZ, time evolution, logarithm
         of density. CDL in red and white, hot post shock layers in light
         blue. Filamentary and dynamic at early times, the CDL becomes
         more 'bulky' later on. Nevertheless, small scale features at
         the surface of the CDL continue to form and vanish. No CDL
         was present at time 0.}
\label{fig:visht}
\end{figure}

\paragraph{Two sided cooling SBIZ:} If both confining shocks cool
efficiently with spatially resolved hot post shock layers (cooling
limit = upstream temperature), the evolution of the CDL at early times
can become much more violent than in the isothermal case
(Fig.~\ref{fig:visht}, left). At later times, the hot post shock zones
rather exert a cushioning effect (Fig.~\ref{fig:visht}, middle and
right). The characteristics of the CDL are hardly affected.
The velocity pdfs show not much change from early to late
times (Fig.~\ref{fig:velopdf}, right).  Much more obvious is the
difference to isothermal simulations, where the velocity field within
the CDL is much more anisotropic (Fig.~\ref{fig:velopdf}, left). Also,
for the same upstream parameters, the root mean square Mach number in
the CDL is much lower in the cooling case than in the isothermal
case~\citep{walder-folini:00, folini-walder:06}. Resolved cooling
layers appear to have a damping effect on the turbulence within the
CDL.
\section{Discussion and Conclusions}
\label{sec:discussion}
Shock bound interaction zones come in many physical varieties and an
intricate interplay exists between the CDL and the confining
shocks. Much work remains to be done, especially also with regard to
observational and theoretical consequences. The presented results 
raise, in particular, the following questions.

One question concerns the wide spread assumption that the relative
speed of high velocity colliding flows translates more or less
directly into the hardness of the X-ray spectrum. At least under
isothermal conditions the confining shocks are not perpendicular to
the colliding flows but steepen with increasing
$M_{\mathrm{u}}$. This should soften the spectrum and shift part of
the emission from the confining shocks to internal shocks of the
CDL~\citep{smith-maclow-heitsch:00}. Whether geometrically thin,
resolved cooling layers behave similarly remains to be studied.

Another topic coming to mind are the relation between SBIZs and 3D
periodic box simulations. The later have enormously improved our
understanding of supersonic turbulence~\citep{maclow:99,
boldyrev-nordlund-padoan:02, padoan-et-al:04, kritsuk-norman:04,
schmidt-et-al:09, federrath-et-al:09}. All the more one may ask to
what degree these results carry over to the CDL in a SBIZ, where
surface effects play a role. Three points may be raised.

In an isothermal SBIZ, $M_{\mathrm{rms}}$,
the energy input into the CDL, and the spatial scale on which
this energy input is modulated all depend on each other.
In 3D box simulations, the energy input and its
spatial modulation - the driving wave length - are usually chosen
independently. The effect of this difference on the turbulence
characteristics remains to be clarified. A second point is the
velocity field, which is clearly anisotropic in the CDL but usually
isotropic by design in 3D box simulations. Finally, when computing a
CDL-averaged structure function (SF) for our 3D CDL (compute SFs for
each point in the CDL, using only points in the CDL, then taking the
average over all SFs) it resembles more closely the observation based
SFs by~\cite{gustafsson-et-al:06} than those typically found for 3D
boxes~\citep{boldyrev-nordlund-padoan:02, padoan-et-al:04,
kritsuk-norman:04}. The reason is not yet clear.  It could be the
finite spatial extension of the CDL, its inhomogeneous, anisotropic
interior, or simply a too coarse resolution.

Further studies of SBIZs are needed, both with basic physics and with
more elaborate models, and with particular attention to the interplay
between the confining shocks and the turbulent CDL.
\acknowledgements 
The authors wish to thank the crew running the HP Superdome at ETH
Zurich and the people of the Swiss Center for Scientific Computing,
CSCS Manno, where the simulations were performed. The authors much
appreciated the stimulating discussions during ASTRONUM, in particular
with E. Audit, P. Hennebelle, A. Kritsuk, and W. Schmidt.

%


\begin{thebibliography}{}

\bibitem[{Anninos} and {Norman}(1996){Anninos} and {Norman}]{anninos-norman:96}
{Anninos}, P. and {Norman}, M.~L. (1996).
\newblock {The role of hydrogen molecules in the radiative cooling and
  fragmentation of cosmological sheets}.
\newblock {\em ApJ\/}, {\bf 460}, 556--568.

\bibitem[{Audit} and {Hennebelle}(2005){Audit} and
  {Hennebelle}]{audit-hennebelle:05}
{Audit}, E. and {Hennebelle}, P. (2005).
\newblock {Thermal condensation in a turbulent atomic hydrogen flow}.
\newblock {\em A\&A\/}, {\bf 433}, 1--13.

\bibitem[{Blondin} and {Marks}(1996){Blondin} and {Marks}]{blondin-marks:96}
{Blondin}, J.~M. and {Marks}, B.~S. (1996).
\newblock {Evolution of cold shock-bounded slabs}.
\newblock {\em New {A}stronomy\/}, {\bf 1}, 235--244.

\bibitem[{Boldyrev} {\em et~al.}(2002){Boldyrev}, {Nordlund}, and
  {Padoan}]{boldyrev-nordlund-padoan:02}
{Boldyrev}, S., {Nordlund}, {\AA}., and {Padoan}, P. (2002).
\newblock {Scaling Relations of Supersonic Turbulence in Star-forming Molecular
  Clouds}.
\newblock {\em ApJ\/}, {\bf 573}, 678--684.

\bibitem[{Federrath} {\em et~al.}(2009){Federrath}, {Klessen}, and
  {Schmidt}]{federrath-et-al:09}
{Federrath}, C., {Klessen}, R.~S., and {Schmidt}, W. (2009).
\newblock {The Fractal Density Structure in Supersonic Isothermal Turbulence:
  Solenoidal Versus Compressive Energy Injection}.
\newblock {\em ApJ\/}, {\bf 692}, 364--374.

\bibitem[{Feldmeier} {\em et~al.}(1997){Feldmeier}, {Puls}, and
  {Pauldrach}]{feldmeier-et-al:97}
{Feldmeier}, A., {Puls}, J., and {Pauldrach}, A.~W.~A. (1997).
\newblock {A possible origin for X-rays from O stars.}
\newblock {\em A\&A\/}, {\bf 322}, 878--895.

\bibitem[{Folini} and {Walder}(2000){Folini} and {Walder}]{folini-walder:00}
{Folini}, D. and {Walder}, R. (2000).
\newblock {Theory of Thermal and Ionization Effects in Colliding Winds of WR+O
  Binaries}.
\newblock In {\em ASP Conf. Ser. 204: Thermal and Ionization Aspects of Flows
  from Hot Stars\/}, pages 267--280.

\bibitem[{Folini} and {Walder}(2006){Folini} and {Walder}]{folini-walder:06}
{Folini}, D. and {Walder}, R. (2006).
\newblock {Supersonic turbulence in shock-bound interaction zones. I. Symmetric
  settings}.
\newblock {\em A\&A\/}, {\bf 459}, 1--19.

\bibitem[{Gustafsson} {\em et~al.}(2006){Gustafsson}, {Brandenburg}, {Lemaire},
  and {Field}]{gustafsson-et-al:06}
{Gustafsson}, M., {Brandenburg}, A., {Lemaire}, J.~L., and {Field}, D. (2006).
\newblock {The nature of turbulence in OMC1 at the scale of star formation:
  observations and simulations}.
\newblock {\em A\&A\/}, {\bf 454}, 815--825.

\bibitem[{Heitsch} {\em et~al.}(2005){Heitsch}, {Burkert}, {Hartmann}, {Slyz},
  and {Devriendt}]{heitsch-et-al:05}
{Heitsch}, F., {Burkert}, A., {Hartmann}, L.~W., {Slyz}, A.~D., and
  {Devriendt}, J.~E.~G. (2005).
\newblock {Formation of Structure in Molecular Clouds: A Case Study}.
\newblock {\em ApJ\/}, {\bf 633}, L113--L116.

\bibitem[{Hennebelle} {\em et~al.}(2008){Hennebelle}, {Banerjee},
  {V{\'a}zquez-Semadeni}, {Klessen}, and {Audit}]{hennebelle-et-al:08}
{Hennebelle}, P., {Banerjee}, R., {V{\'a}zquez-Semadeni}, E., {Klessen}, R.~S.,
  and {Audit}, E. (2008).
\newblock {From the warm magnetized atomic medium to molecular clouds}.
\newblock {\em A\&A\/}, {\bf 486}, L43--L46.

\bibitem[{Hunter} {\em et~al.}(1986){Hunter}, {Sandford}, {Whitaker}, and
  {Klein}]{hunter-et-al:86}
{Hunter}, J.~H., {Sandford}, M.~T., {Whitaker}, R.~W., and {Klein}, R.~I.
  (1986).
\newblock {Star formation in colliding gas flows}.
\newblock {\em ApJ\/}, {\bf 305}, 309--332.

\bibitem[{Kang} {\em et~al.}(2005){Kang}, {Ryu}, {Cen}, and
  {Song}]{kang-et-al:05}
{Kang}, H., {Ryu}, D., {Cen}, R., and {Song}, D. (2005).
\newblock {Shock-heated Gas in the Large-Scale Structure of the Universe}.
\newblock {\em \apj\/}, {\bf 620}, 21--30.

\bibitem[{Kritsuk} and {Norman}(2004){Kritsuk} and {Norman}]{kritsuk-norman:04}
{Kritsuk}, A.~G. and {Norman}, M.~L. (2004).
\newblock {Scaling Relations for Turbulence in the Multiphase Interstellar
  Medium}.
\newblock {\em ApJ\/}, {\bf 601}, L55--L58.

\bibitem[{Mac Low}(1999){Mac Low}]{maclow:99}
{Mac Low}, M.-M. (1999).
\newblock {The Energy Dissipation Rate of Supersonic, Magnetohydrodynamic
  Turbulence in Molecular Clouds}.
\newblock {\em ApJ\/}, {\bf 524}, 169--178.

\bibitem[{Myasnikov} and {Zhekov}(1998){Myasnikov} and
  {Zhekov}]{myasnikov-zhekov:98}
{Myasnikov}, A.~V. and {Zhekov}, S.~A. (1998).
\newblock {Dissipative models of colliding stellar winds - I. Effects of
  thermal conduction in wide binary systems}.
\newblock {\em MNRAS\/}, {\bf 300}, 686--694.

\bibitem[{Nussbaumer} and {Walder}(1993){Nussbaumer} and
  {Walder}]{nussbaumer-walder:93}
{Nussbaumer}, H. and {Walder}, R. (1993).
\newblock {Modification of the nebular environment in symbiotic systems due to
  colliding winds}.
\newblock {\em A\&A\/}, {\bf 278}, 209--225.

\bibitem[{Owocki} {\em et~al.}(1988){Owocki}, {Castor}, and
  {Rybicki}]{owocki-et-al:88}
{Owocki}, S.~P., {Castor}, J.~I., and {Rybicki}, G.~B. (1988).
\newblock {Time-dependent models of radiatively driven stellar winds. I -
  Nonlinear evolution of instabilities for a pure absorption model}.
\newblock {\em ApJ\/}, {\bf 335}, 914--930.

\bibitem[{Padoan} {\em et~al.}(1997){Padoan}, {Nordlund}, and
  {Jones}]{padoan-nordlund-jones:97}
{Padoan}, P., {Nordlund}, A., and {Jones}, B.~J.~T. (1997).
\newblock {The universality of the stellar initial mass function}.
\newblock {\em MNRAS\/}, {\bf 288}, 145--152.

\bibitem[{Padoan} {\em et~al.}(2004){Padoan}, {Jimenez}, {Nordlund}, and
  {Boldyrev}]{padoan-et-al:04}
{Padoan}, P., {Jimenez}, R., {Nordlund}, {\AA}., and {Boldyrev}, S. (2004).
\newblock {Structure Function Scaling in Compressible Super-Alfv{\'e}nic MHD
  Turbulence}.
\newblock {\em Physical Review Letters\/}, {\bf 92}(19), 191102--+.

\bibitem[{Passot} and {V{\'a}zquez-Semadeni}(1998){Passot} and
  {V{\'a}zquez-Semadeni}]{passot-vazquez:98}
{Passot}, T. and {V{\'a}zquez-Semadeni}, E. (1998).
\newblock {Density probability distribution in one-dimensional polytropic gas
  dynamics}.
\newblock {\em Physical Review Letters\/}, {\bf 58}, 4501--4510.

\bibitem[{Schmidt} {\em et~al.}(2009){Schmidt}, {Federrath}, {Hupp}, {Kern},
  and {Niemeyer}]{schmidt-et-al:09}
{Schmidt}, W., {Federrath}, C., {Hupp}, M., {Kern}, S., and {Niemeyer}, J.~C.
  (2009).
\newblock {Numerical simulations of compressively driven interstellar
  turbulence. I. Isothermal gas}.
\newblock {\em A\&A\/}, {\bf 494}, 127--145.

\bibitem[{Smith} {\em et~al.}(2000){Smith}, {Mac Low}, and
  {Heitsch}]{smith-maclow-heitsch:00}
{Smith}, M.~D., {Mac Low}, M.-M., and {Heitsch}, F. (2000).
\newblock {The distribution of shock waves in driven supersonic turbulence}.
\newblock {\em A\&A\/}, {\bf 362}, 333--341.

\bibitem[{Stevens} {\em et~al.}(1992){Stevens}, {Blondin}, and
  {Pollock}]{stevens-et-al:92}
{Stevens}, I.~R., {Blondin}, J.~M., and {Pollock}, A.~M.~T. (1992).
\newblock {Colliding winds from early-type stars in binary systems}.
\newblock {\em ApJ\/}, {\bf 386}, 265--287.

\bibitem[{Strickland} and {Blondin}(1995){Strickland} and
  {Blondin}]{strickland-blondin:95}
{Strickland}, D. and {Blondin}, J.~M. (1995).
\newblock {Numerical analysis of the dynamic stability of radiative shocks}.
\newblock {\em ApJ\/}, {\bf 449}, 727--738.

\bibitem[{V{\'a}zquez-Semadeni} {\em et~al.}(2006){V{\'a}zquez-Semadeni},
  {Ryu}, {Passot}, {Gonz{\'a}lez}, and {Gazol}]{vazquez-semadeni-et-al:06}
{V{\'a}zquez-Semadeni}, E., {Ryu}, D., {Passot}, T., {Gonz{\'a}lez}, R.~F., and
  {Gazol}, A. (2006).
\newblock {Molecular Cloud Evolution. I. Molecular Cloud and Thin Cold Neutral
  Medium Sheet Formation}.
\newblock {\em ApJ\/}, {\bf 643}, 245--259.

\bibitem[{Vishniac}(1994){Vishniac}]{vishniac:94}
{Vishniac}, E.~T. (1994).
\newblock {Nonlinear instabilities in shock-bounded slabs}.
\newblock {\em ApJ\/}, {\bf 428}, 186--208.

\bibitem[{Walder} and {Folini}(1996){Walder} and {Folini}]{walder-folini:96}
{Walder}, R. and {Folini}, D. (1996).
\newblock {Radiative cooling instability in 1D colliding flows.}
\newblock {\em A\&A\/}, {\bf 315}, 265--283.

\bibitem[{Walder} and {Folini}(1998){Walder} and {Folini}]{walder-folini:98}
{Walder}, R. and {Folini}, D. (1998).
\newblock {Knots, filaments, and turbulence in radiative shocks}.
\newblock {\em A\&A\/}, {\bf 330}, L21--L24.

\bibitem[{Walder} and {Folini}(2000a){Walder} and {Folini}]{amaze:00}
{Walder}, R. and {Folini}, D. (2000a).
\newblock {A-MAZE: A code package to compute 3D magnetic flows, 3D NLTE
  radiative transfer, and synthetic spectra}.
\newblock In H.~J. G. L.~M. Lamers and A.~Sapar, editors, {\em Thermal and
  Ionization Aspects of Flows from Hot Stars: Observations and Theory\/}, ASP
  Conference Series, pages 281--285.

\bibitem[{Walder} and {Folini}(2000b){Walder} and {Folini}]{walder-folini:00}
{Walder}, R. and {Folini}, D. (2000b).
\newblock {On the Stability of Colliding Flows: Radiative Shocks, Thin Shells,
  and Supersonic Turbulence}.
\newblock {\em Ap\&SS\/}, {\bf 274}, 343--352.

\end{thebibliography}
\end{document}